\documentclass[aps,prb,preprint]{revtex4-2}

\usepackage{amsmath,amssymb}
\usepackage{graphicx}
\usepackage{epsfig}
\usepackage{multirow}
\usepackage{color}

\newcommand{\vect}[1]{\boldsymbol{#1}}
\newcommand{\bhat}[1]{\hat{\boldsymbol{#1}}}

\newcommand{\zhat}{\bhat{z}}

\begin{document}

\input{epsf}

\title{Modeling the induced voltage of a disk magnet in free fall}
\author{Nolan Samboy}

\affiliation{Department of Physical and Biological Sciences,\\ Western New England University,\\
            1215 Wilbraham Road, Springfield, MA 01119}

\date{\today}
\begin{abstract}
We drop a circular disk magnet through a thin coil of wire, record the induced voltage,
and compare the results 
to an analytic model based on the dipole approximation
and Faraday's law, which predicts that the difference between the voltage peak magnitudes corresponding
to the entry and exit of the magnet should be in proportion to
$z_0^{-1/2}$, where $z_0$ is the initial height of the magnet above the center of the coil.
Agreement between the model and experimental data are excellent. This easily-reproduced experiment provides an opportunity for students at a range of
levels to quantitatively explore the effects of magnetic induction.
\end{abstract}

\maketitle
\section{Introduction}
\label{sec:intro}
Faraday's law of electromagnetic induction is fundamental to any introductory physics sequence. While it is 
straightforward to demonstrate qualitatively, providing students with a quantitative experiment is more challenging. 
A common approach is to drop a magnet through a conducting
loop and measure consequences of that relative motion. These experiments tend to fall into two basic 
categories: (i) inducing resistive magnetic forces via eddy currents,~\cite{Derby-drag, Hahn-drag, Levin-drag,
Roy-drag, Irvine-drag}
and (ii) inducing a voltage signal that is visualized in real time using computer acquisition 
software.~\cite{Nicklin, Kingman, Amrani,Reeder, Gadre}
In this paper, we follow the second group and describe an experiment where students drop a 
circular disk magnet through an induction coil and relate the resulting
peak voltage values to the drop height of the magnet. Such an approach presents a good mix of intuition
gained from introductory mechanics and electromagnetic phenomena. The recent paper of Gadre et al. describes
an experiment similar to that which we describe here, but our analysis is more thorough and explicitly compares
experimental results to theoretical predictions. \cite{Gadre}
While some of the details provided here are outside the scope of an introductory-level course, 
they would be accessible to upper-level students, and are critical to the discussion contained herein.

\section{Free-falling magnet}
\label{sec:freefall}
\begin{figure}[h]
	\centering
		\includegraphics[width=2.0in]{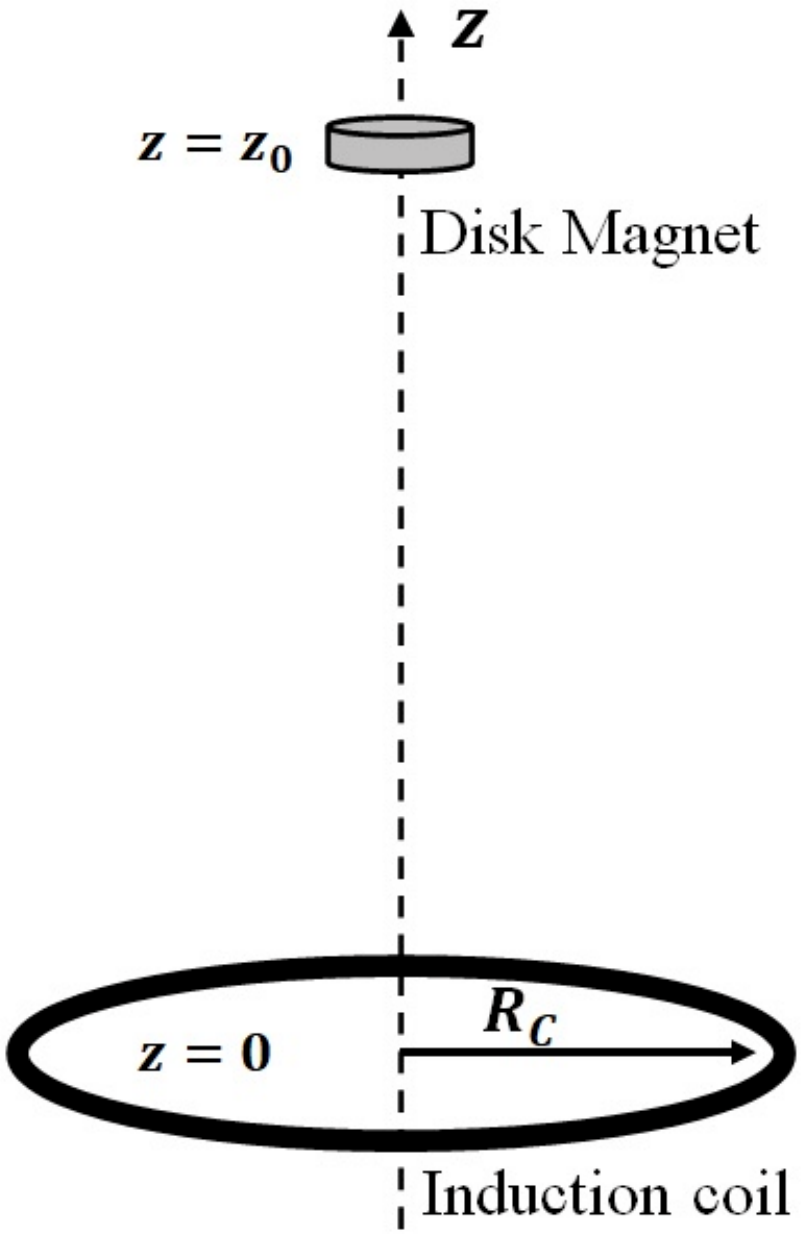}
	\caption{Magnet-coil geometry. The magnet and coil share a common $z$-axis; 
					 $z=0$ corresponds to the center of the induction coil,
					 and $z=z_0$ is the starting height of the magnet.}
	\label{fig:magnetfall}
\end{figure}

In this section we provide an analytic model for the induced voltage signal produced by 
a disk magnet falling freely through an open coil of radius $R_c$ and negligible thickness. 
As shown in Fig.~\ref{fig:magnetfall}, the starting
height of the magnet $z_0$ is defined as the distance from the mid-point of the coil to the mid-point
of the magnet, 
which is assumed to fall freely through the center of the coil along the $z$-axis. 
We ignore air drag as well as any resistive magnetic forces that act on the magnet from the coil.
\subsection{Modeling the disk magnet}
\label{subs:model}
We assume that a circular disk magnet can be modeled as an ideal $N$-turn circular loop of current $I$,
with dimensions equal to that of the magnet, save for the thickness, which we assume to be negligible. 
As the magnet falls along the $z$-axis through the induction coil, only the $z$-component of the magnetic field
will contribute to the flux, and thus to the induced voltage. Therefore, we only need to concern ourselves with 
the $z$-component of the magnetic field, which can be found using the Biot-Savart law
and the standard dipole approximation:~\cite{Jackson}
\begin{equation}
\label{eq:circapp}
\vect{B_z}(\vect{r})\approx\frac{kR^2}{4}\frac{(2z^2-\rho^2)}{(\rho^2+z^2)^{5/2}}\,\zhat\,\,.
\end{equation}
Here, $R$ is the radius of the magnet, $\rho = \sqrt{x^2+y^2}$ is the cylindrical coordinate of the field
point, and we have set $k=N\mu_0I$, where $k$ is an empirical constant to be determined.
The total magnetic flux (linkage) of such a field through the open area of an induction coil with radius
$R_c$ and number of turns $N_c$ is found via direct integration to be:
\begin{equation}
\label{eq:totalflux}
\Phi_{\rm tot}(z) = \frac{N_c\,k\pi R^2}{2}\frac{R_c^2}{(R_c^2+z^2)^{3/2}}\,\,.
\end{equation}
A full derivation of Eqs.~\eqref{eq:circapp} and Eq.~\eqref{eq:totalflux} appears in the Supplementary Material.~\cite{supplemental}
\subsection{Modeling the Voltage Signal}
\label{subs:signal}
For a magnet that is moving, $z\equiv z(t)$ and so Eq.~\eqref{eq:totalflux} becomes a function of time where the flux
increases as the magnet approaches the coil, reaches a maximum when it is at the center of the coil, and
then decreases as it exits the coil.
The resulting voltage signal in the coil is found from Faraday's law of induction and the chain rule to 
be:
\begin{equation}
\label{eq:faradaychain}
V = -\frac{d\Phi_{\rm tot}(z)}{dt} = -\frac{d\Phi_{\rm tot}(z)}{dz}\,\frac{dz}{dt} 
     = \frac{3 N_c\,k\pi R^2 R_c^2\,z}{2(R_c^2+z^2)^{5/2}}\,v \,\,,
\end{equation}
where $v$ is the speed of the moving magnet. 
When the speed is constant, Eq.~\eqref{eq:faradaychain} is an 
anti-symmetric function:
the two voltage peaks occur at $z_{p,1}=+R_c/2$ and $z_{p,2}=-R_c/2$ ($z=0$ at the center of the coil),
and the peaks are equal in value but opposite in polarity. 
This was extensively verified in Kingman, et. al.~\cite{Kingman}

For an accelerating magnet, Eq.~\eqref{eq:faradaychain} is no longer exactly anti-symmetric. Since the speed of
the magnet is higher as it exits the coil than when it enters, we should expect
the later peak voltage $|V_{p,2}|$ to be larger than the earlier one $|V_{p,1}|$. 
Further, the voltage peaks do not occur exactly at $z_{p,i}=\pm R_c/2$; however, the shifts away
from these points are small enough to ignore (see Appendix~\ref{appB}).

Since measuring the instantaneous speed of an accelerating (free falling) magnet is difficult, we instead seek a relationship
between the initial height of the magnet $z_0$ and the measured peak voltages $V_{p,i}$ of the
signal. For a magnet in free fall, its speed at any position $z$ is related to the drop height $z_0$ through 
kinematics: $v = \sqrt{2g(z_0-z)}$. Inserting this into Eq.~\eqref{eq:faradaychain} and assuming the peak voltages
$V_{p,i}$ occur at the specific positions $z_{p,i}$, we find the peak voltage values to be:
\begin{equation}
\label{eq:Vpeaks-z}
V_{p,i}=\frac{3 N_c\,k\pi R^2 R_c^2\,z_{p,i}}{2(R_c^2+z_{p,i}^2)^{5/2}}\,\sqrt{2g(z_0-z_{p,i})}\,\,.
\end{equation}
If we further assume that the peak positions occur approximately at $z_{p,1} = +R_c/2$ and $z_{p,2} = -R_c/2$
and let the drop height be much larger than these positions (i.e. $z_0 >> R_c/2$), we can expand the square root 
and find:
\begin{align}
\label{eq:Vpeak1}
|V_{p,1} (z_0)| &= \frac{3}{4}\left(\frac{4}{5}\right)^{5/2}
\frac{k N_c\pi R^2 g}{R_c^2}\left(\sqrt{\frac{2z_0}{g}}-\frac{R_c}{2\sqrt{2gz_0}} \right) \\
\label{eq:Vpeak2}
|V_{p,2} (z_0)| &= \frac{3}{4}\left(\frac{4}{5}\right)^{5/2}
\frac{k N_c\pi R^2 g}{R_c^2}\left(\sqrt{\frac{2z_0}{g}}+\frac{R_c}{2\sqrt{2gz_0}} \right)\,\,.
\end{align}
Here, $V_{p,1} (z_0)$ is the first (earlier) peak voltage and $V_{p,2} (z_0)$ 
is the second (later) peak voltage. We express the results
as magnitudes since the polarity depends on the experimental set up.
We note that the difference between the magnitudes of the two peaks asymptotically approaches zero with increasing
drop height:
\begin{equation}
\label{eq:Vpeakdiff}
|V_{p,2} (z_0)|-|V_{p,1} (z_0)| = \frac{3}{4}\left(\frac{4}{5}\right)^{5/2}
\frac{k N_c\pi R^2}{R_c}\sqrt{\frac{g}{2}}\,\,\frac{1}{\sqrt{z_0}}\,\,.
\end{equation} 
While upper-level E\&M students should be capable of applying the binomial expansion, a more qualitative
argument for the $z_0^{-1/2}$-dependence can be made for introductory students by noting that
the difference in the peak voltage values is proportional to the
difference in the corresponding speed values, i.e. $\Delta V_p\propto\Delta v$. Here, $\Delta v=g\Delta t$, where
$\Delta t$ is the time between the peaks.
If the drop height is sufficiently large, then $\Delta t$ will be much smaller than the total drop time and so the
magnet's change in speed will be minimal during this $\Delta t$ interval.
Approximating the speed as an average value during this time
yields $\Delta v = g\dfrac{\Delta y}{v_{\rm ave}}$, where $\Delta y$ is the vertical
distance traveled during $\Delta t$ and
$v_{\rm ave}\approx\sqrt{2gz_0}$, which is the magnet's speed at the center of the coil. 
Thus, we arrive at $\Delta V\propto z_0^{-1/2}$, which is straightforward to verify experimentally.
\section{Experiment}
\label{sec:experiment}
\begin{figure}[h]
  \centering
	  \includegraphics[width=3in]{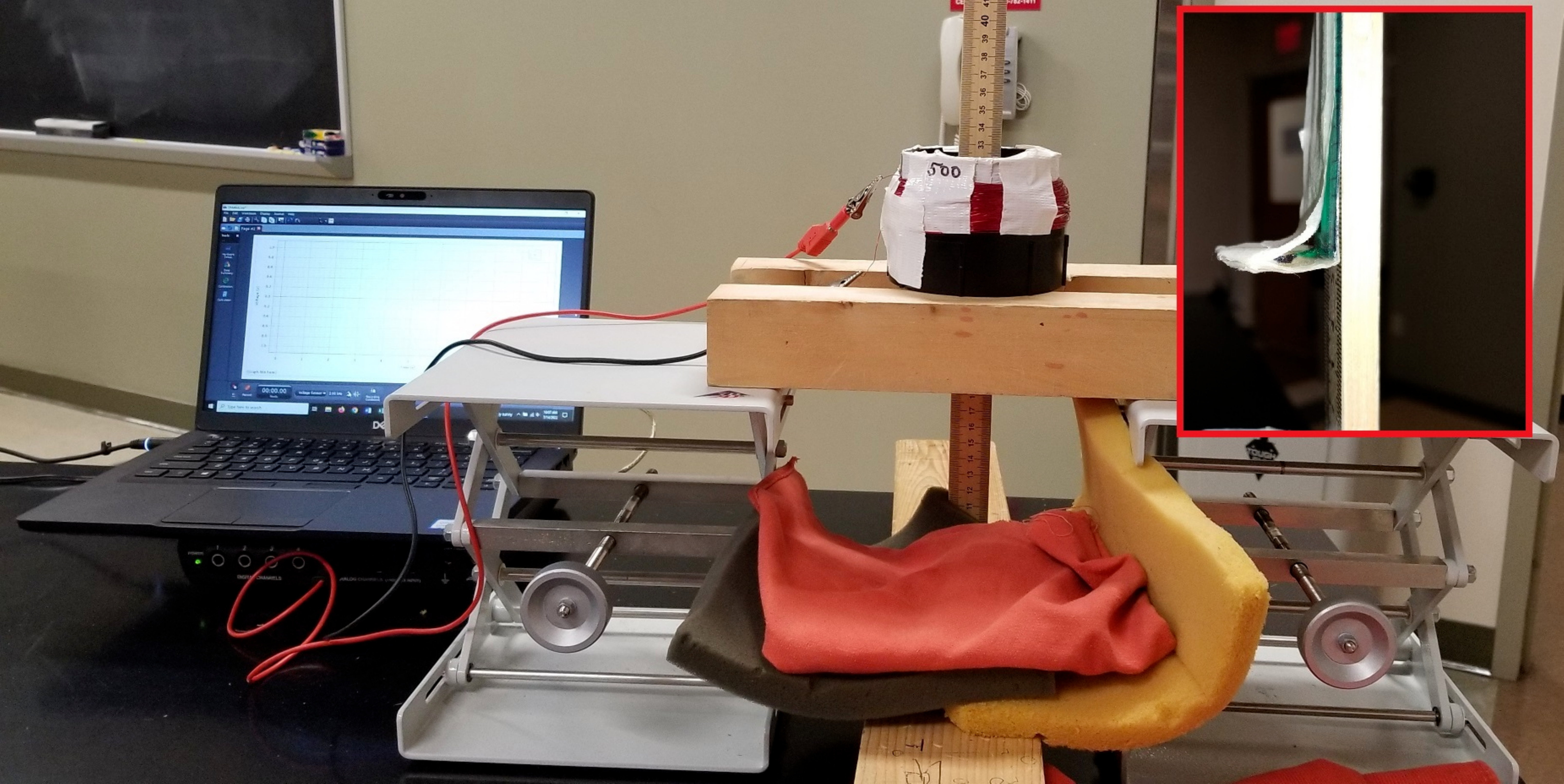}
\caption{Setup of the experiment: A 500-turn induction coil is elevated by two jack stands, while a meter stick is positioned
         inside the coil to measure the drop height of the magnet; the induced voltage signal is measured by a Pasco
				 750 Computer Interface in conjunction with Capstone software. Inset: The disk magnet is taped to a plastic
				 ruler to allow its dipole moment to stay aligned with the coil's symmetry axis as it falls.
}
\label{fig:setup}
\end{figure}

A disk magnet of radius $R=0.953$ cm and thickness $t=0.318$ cm was taped to a plastic ruler to 
facilitate consistent dropping through a $N_c=500$-turn coil of copper wire wrapped around
a piece of ABS plastic piping. The coil, of average
radius  $R_c=4.83$ cm and thickness $h=2.42$ cm,
was set on two blocks of wood and  elevated by two jack stands;
foam padding was set beneath the coil to protect the magnet as it fell, and a meter stick was stationed inside 
the coil to measure the drop height; see Fig.~\ref{fig:setup}.
While it might not be obvious that the dipole approximation  
should apply here given the relative dimensions of the magnet and the coil, we will see below that it does yield
a very good fit with the experimental data.

To determine the $k$-value of the disk magnet, a FW Bell model 4048 Gauss/Tesla meter ($\pm\,2\%$ tolerance)
was used to measure the magnetic field $B_z$ along the main axis from 4.0~cm to 10.5~cm in 0.5~cm intervals.
We plotted $B_z$ vs. $(R^2+z^2)^{-3/2}$
and found $k=0.00425$~T$\cdot$m from the slope of the linear fit.~\cite{supplemental}
This corresponds to a magnetic dipole moment of $m= 0.965$ A$\cdot$m$^2$.
\subsection{Data Acquisition}
\label{subs:data}
The voltage measurements were made with a Pasco 750 Computer Interface in conjunction with Pasco Capstone 
software. As the magnet fell through the main axis of the coil, the software displayed the induced 
voltage signal as a function of time; a 2,000 Hz sampling rate was used for every trial. 
The starting height of the magnet was varied from 1.0 cm to 5.0 cm in 1.0 cm intervals, and then
5.0~cm to 45~cm in 2.5~cm intervals. All values are measured from the center of the coil $z=0$.
For each starting height, the magnet was dropped five times and the two peak voltages were recorded for every
trial; all results are thus presented as an average (data point) and an uncertainty (error bars). 
\subsection{Results}
\label{subs:results}
\begin{figure}[h]
	\centering
		\includegraphics[width=3in]{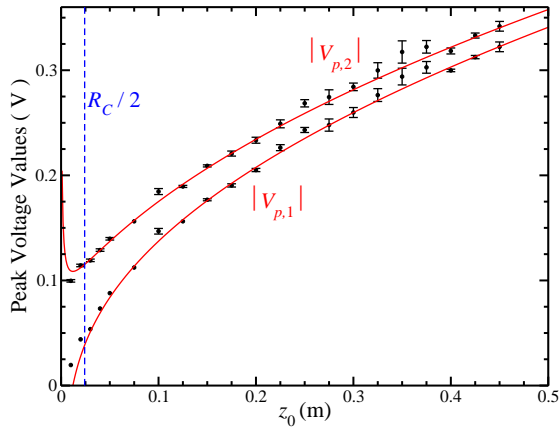}
	\caption{(Color online) Plots of peak voltage magnitudes against the drop height of the magnet; both plots are
	          presented as experimental data (black data points/error bars) along with the corresponding analytic model 
						(solid red line).
						The theoretical curves
						are calculated by directly applying the experimental parameters to Eq.~\eqref{eq:Vpeak1} (lower curve)
						and Eq.~\eqref{eq:Vpeak2} (upper curve).
						The model is valid for drop heights larger than half the radius of the induction coil $R_c/2$, indicated
						by a blue dashed line.
						We note the agreement between the predicted curves and the empirical data for $z_0 > R_c/2$.
						}
	\label{fig:peak-data}
\end{figure}

\begin{figure}[h]
	\centering
		\includegraphics[width=3in]{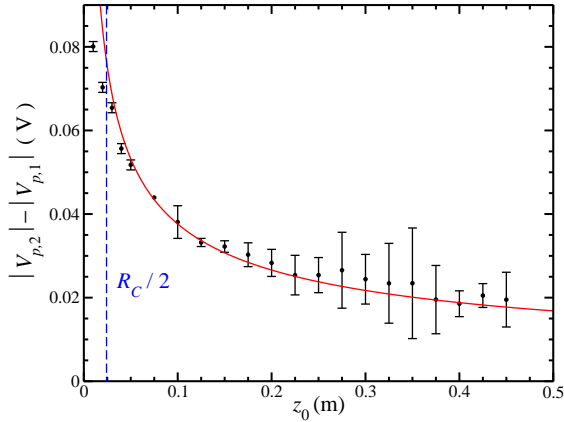}
	\caption{(Color online) Difference in the two voltage peaks plotted against the drop height. The data points and
	         error bars correspond to the experimental results, while the solid red line 
	         corresponds to Eq.~\eqref{eq:Vpeakdiff}, with all relevant experimental parameters applied.
					 The model is valid for drop heights larger than half the radius of the induction coil
					 $R_c/2$, marked by a dashed blue line.
					 The predicted $z_0^{-1/2}$-dependence is confirmed, given the agreement between theory and data for
					 $z_0 > R_c/2$.
					}
	\label{fig:Delta-peaks}
\end{figure}

In Fig.~\ref{fig:peak-data} we plot the experimental peak voltage values against the drop height,
along with the theoretical predictions given by Eq.~\eqref{eq:Vpeak1} and Eq.~\eqref{eq:Vpeak2}. 
The uncertainties are given by the sample standard deviation of the mean
of the measurements, while the the theoretical curves were obtained directly
from Eqs.~\eqref{eq:Vpeak1} and~\eqref{eq:Vpeak2}, 
along with the experimental parameters defined earlier. We find very good agreement between the experimental
data and the theoretical models for drop heights greater than $R_c/2\sim$~2.4 cm,
which was an assumed condition in the derivation of Eqs.~\eqref{eq:Vpeak1} and~\eqref{eq:Vpeak2}.
We mark the position $z_0 = R_c/2$ explicitly in Fig.~\ref{fig:peak-data} and Fig.~\ref{fig:Delta-peaks}, 
and note that the model begins to break down for drop heights less than this value.

The upper limit of the data set was mainly a practical one: consistently dropping the magnet through the center
of the induction coil  became too difficult for drop heights larger than $\sim$ 45 cm.
While drag effects could also begin to 
manifest at larger drop heights, we show later that these effects are minimal for the magnet used in this experiment.

As mentioned earlier, the $z_0^{-1/2}$-dependence is key to accurately modeling the peak voltage magnitudes. 
We demonstrate this explicitly in  Fig.~\ref{fig:Delta-peaks}
by plotting the difference between the peak voltages against the drop
height. The uncertainties are found via propagation of the error values given in 
Fig.~\ref{fig:peak-data}.
We again find the theoretical curve by applying 
the relevant experimental parameters directly to Eq.~\eqref{eq:Vpeakdiff}, and find very good agreement
for all drop heights larger than 2.4 cm.

\begin{figure}[h]
	\centering
		\includegraphics[width=3in]{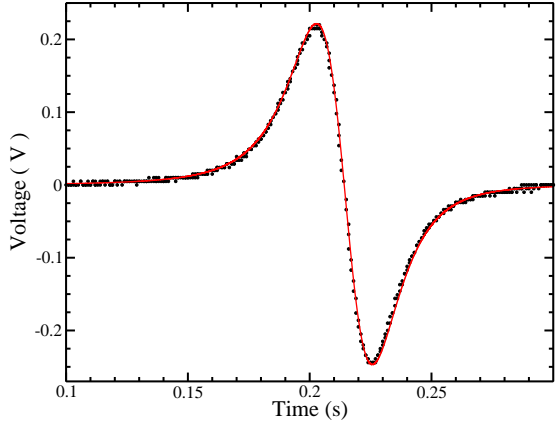}
	\caption{(Color online) Representative trial run ($z_0 = 30$ cm) of experimental voltage data (black points) 
	         and theoretical line shape (red solid line) as functions of time. The data set has been horizontally shifted
					 to align with the time scale defined by the theoretical calculation ($t=0$ at release),
					 such that it crosses the $V=0$ axis at the same time $t\approx 0.21$s as the theoretical curve.
					 We note the excellent agreement between the data and the model.
					}
	\label{fig:voltage}
\end{figure}

Finally, if we define Eq.~\eqref{eq:faradaychain} explicitly as a function of time by letting $z(t) = z_0-\frac{1}{2}gt^2$
and $v(t) = -gt$, we arrive at the following expression which can be numerically evaluated:
\begin{equation}
\label{eq:emf}
V(t) = \frac{3gt N_c\,k\pi R^2 R_c^2\,(z_0-\frac{1}{2}gt^2)}{2(R_c^2+(z_0-\frac{1}{2}gt^2)^2)^{5/2}}\,\,.
\end{equation}
In Fig.~\ref{fig:voltage} we match the complete theoretical line shape
as predicted by Eq.~\eqref{eq:emf} to the induced voltage signal for a representative sample run ($z_0 = 30$ cm);
the agreement between the experimental signal and the theoretical model is essentially perfect. This further
validates the analytical approach taken in Section~\ref{sec:freefall}.
\subsection{Error and Limitations}
For the experiment described here, the largest source of experimental error comes from dropping the 
magnet by hand through the center of the induction coil from the same height for any five repeated trials.
Anecdotally, it was observed that if the magnet drifted away from the symmetry axis as it fell, the voltage peaks were
affected by this motion. The data indicates that this was more of an issue at larger drop heights, 
where the average peak voltage values tend to skew larger than the theoretical line shapes, 
in particular for the later peak. While numerical modeling seems to suggest that a falling
magnet with non-zero horizontal motion can result in a larger flux change (and thus larger voltage peaks), 
we cannot conclusively verify this yet. Overall, however, these effects  seem to be random, and they
are small: differences between the data points and the predicted curves are on the order
of millivolts, which begins to approach the tolerance of the instrumentation.

For larger drop heights, it is also reasonable to suspect that air drag would begin to affect the magnet's speed
and thus affect the voltage measurements. We modeled a quadratic drag force on the square 2.30~cm~$\times~$2.30~cm
cross section of masking tape holding the magnet to the ruler (total mass of 0.02 kg), and found that air
drag accounted for a reduction of only 0.6\% of the speed at the largest drop height used in this experiment.

Similarly, resistive drag effects due to the induced currents in the coil were also minimal. Since air resistance
is not a significant factor, the magnet's kinetic energy at the center of the coil should be equal to its 
potential energy at release $mgz_0$, which is 88 mJ for the largest drop height used in this experiment. The energy lost due to
inducing a current in the coil is found by integrating the electric power over time:
\begin{equation}
U = \int{P(t)\,dt} = \int_0^{\infty}{\frac{V^2(t)}{r}\,dt}\,\,
\end{equation}
where $V(t)$ is defined in Eq.~\eqref{eq:emf} and $r$ is the resistance of the coil, measured to be 21.7~$\Omega$.
This calculation yields an energy loss of 0.123 mJ for the largest drop height, 
roughly $0.1\%$ of the initial energy. Repeating for the smallest drop height yielded an initial energy
value of 1.96 mJ and an energy loss of 0.0188 mJ, or about $1\%$.

The theoretical model hinges on two main assumptions. One is that the drop
height of the magnet is much larger than half the coil radius. This allows us to define the locations
of the voltage extrema at $z_{p,i}=\pm R_c/2$ as well as to employ the binomial expansion on the speed term in the peak
voltage definitions. This assumption also coincides with the range of validity for the dipole
approximation.

The second assumption is that both the magnet and the induction coil be treated as ``thin rings'' of 
negligible thickness. For a cylindrical object, this approximation is gauged by the ratio of the cylinder's
length $L$ to its diameter $D$; as long as this ratio is ``small enough,'' the approximation is valid. 
Determining exactly where this approximation breaks down depends on the details of the individual
experiment, but for the objects
used here, the magnet had an $L:D$ ratio of about $1:6$, while the induction coil had a ratio of about $1:4$.
Based on the results of our experiment, we conclude that these ratios are indeed ``small enough'' for the models
to be considered valid. Separate numerical modeling (not shown) indicates that the thin ring model for the induction coil
begins to noticeably separate from the solenoid model around $L=D/2,$ which would correspond to the edges of the 
coil being located at the assumed peak
voltage positions $z_{p,i}=\pm R_c/2$. While this should not necessarily be considered \emph{the point} at which the 
model breaks down, it could serve as a rough upper limit when designing the experiment.
\section{Conclusion}
\label{sec:conc}
In this paper, we have developed an analytic model of the induced voltage signal
produced by a disk magnet falling freely through a thin induction coil. The initial motivation for this work was
to develop a quantitative experiment to test Faraday's law; however, we were pleasantly surprised at the 
robust opportunities for engagement of students at various course levels that this investigation provides.
All of the measurements are easily obtained in an introductory setting, requiring materials and equipment
either generally available or easy to obtain.

While a similar but slightly more rigorous derivation of the peak
voltage values can be done in the time coordinate, the argument presented here in the $z$-coordinate
allows students to make an intuitive connection between introductory kinematics and the observed voltage
signals. Students can be led through a series of questions and/or asked to make their own speculations 
about what they expect to see based on their knowledge of free fall. 
At the introductory level, students could then simply measure the drop height of the magnet $z_0$, the two 
resulting voltage peak values $|V_{p,1}|$, $|V_{p,2}|$, and test their predictions. It would also be straightforward
at that point to verify the $z_0^{-1/2}$-dependence of $|V_{p,2}|-|V_{p,1}|$
(as in Fig.~\ref{fig:Delta-peaks}). 
Depending on the expectation of the level of 
rigor for the experiment, students could also measure the experimental parameters 
required to plot the theoretical curve for direct comparison to the data set. 

In an upper-level course, students could be expected to replicate some or all of the derivations contained in this
paper~\cite{supplemental}, have a stronger understanding of the model, and be able to explore the
consequences of varying different parameters on the resulting voltage signal.
For example, magnets of different radii or dipole moments could be used, along with different 
induction coils of varying radii or number of turns. For comparison to the experimental voltage signals,
theoretical line shapes are easily produced using any graphing software or numerical coding platforms.
While the application of the dipole approximation makes the full model accessible to upper-level
students, it is also conceivable that it could be eschewed in favor of a complete elliptic integral
approach that could be investigated as part of an advanced project; this could be done
while experimenting with an induction coil that is similar in size to the magnet.

Finally, the work included here could be extended to examine a disk 
magnet moving through a solenoid induction coil rather than a thin-ring coil, or a bar magnet moving through
various coil geometries. It could also be interesting to evaluate the resulting voltage signals of a disk magnet 
moving with varied accelerations as controlled via an Atwood's machine.
\appendix
\section{Approximating the Peak Voltage Positions}
\label{appB}
Beginning with Eq.~\eqref{eq:faradaychain} and applying the free-fall condition for $v(z)$, we arrive at the following 
general expression for the voltage signal as a function of position $z$:
\begin{equation}
\label{eq:emf-z}
V(z) = \frac{3 N_c\,k\pi R^2 R_c^2\,z}{2(R_c^2+z^2)^{5/2}}\,\sqrt{2g(z_0-z)} \,\,.
\end{equation}
The voltage peaks for this signal occur at the peak position $z=z_p$ (assuming $z=0$ at the center of the coil), where
the derivative of Eq.~\eqref{eq:emf-z} is equal to zero.
Taking the derivative with respect to $z$, setting equal to zero, and simplifying yields:
\begin{equation}
\label{eq:demf-z}
\frac{d V(z)}{dz}\bigg\rvert_{z=z_p}
= 7z_p^3-8z_0z_p^2-3R_c^2 z_p + 2R_c^2z_0 = 0\,\,.
\end{equation}
While these $z_p$ values
can be solved for numerically, the results are not particularly instructive. Instead,
we can reasonably approximate values for $z_p$ by assuming that the drop height $z_0$ is much larger than $z_p$.
Factoring out a $z_0^3$ gives
\begin{equation}
\label{eq:z0-factor}
z_0^3\left[7\left(\frac{z_p}{z_0}\right)^3 - 8\left(\frac{z_p}{z_0}\right)^2 - 3\,\frac{R_c^2z_p}{z_0^3}+2\,\frac{R_c^2}{z_0^2}\right] = 0\,\,,
\end{equation}
Ignoring the cubic term, we can solve the remaining quadratic for $z_p$:
\begin{equation}
\label{eq:zp}
z_p = -\frac{3R_c^2\pm\sqrt{9R_c^4+64R_c^2z_0^2}}{16 z_0}\,\,,
\end{equation}
which we can further simplify by applying the condition that $z_0~>>~R_c/2$:
\begin{equation}
\label{eq:zp2}
z_p \approx -\frac{3R_c^2}{16 z_0}\pm \frac{R_c}{2} 
\approx\pm\frac{R_c}{2}\,\,. 
\end{equation}
Thus, if the drop height is (much) larger than half the radius of the induction coil, the voltage peaks will
occur approximately at $\pm R_c/2$, which is where they peak in the constant-speed case.
\begin{acknowledgments}
The author gratefully acknowledges Joseph Gallant for carefully proof-reading early versions of this manuscript, Karl Martini
and Sam Emery for discussions of uncertainty analysis, and the reviewers for their thoughtful comments, helpful suggestions, and overall insight.
\end{acknowledgments}
The author has no conflicts to disclose.
%
\end{document}